\documentclass[12pt,final,5p,times,twocolumn,]{elsarticle}

\usepackage{amssymb}
\usepackage{amsmath}
\usepackage{color}
\usepackage{bm}
\usepackage{booktabs}
\usepackage{caption}
\usepackage{stfloats}
\usepackage[squaren]{SIunits}
\biboptions{numbers,sort&compress}

\journal{Materials today}

\begin{document}

\begin{frontmatter}



\title{Optimal isotropic, reusable truss lattice material with near-zero Poisson's ratio}


\author{Xueyan Chen$^{a,b}$, Johnny Moughames$^{b}$, Qingxiang Ji$^{a,b}$, Julio Andr\'es Iglesias Mart\'inez$^{b}$, Huifeng Tan$^{a,*}$, Samia Adrar$^{b}$, Nicolas Laforge$^{b}$, Jean-Marc Cote$^{b}$, S\'ebastien Euphrasie$^{b}$, Gwenn Ulliac$^{b}$, Muamer Kadic$^{b}$, and Vincent Laude$^{b}$}

\address{%
$^{a}$ \quad National Key Laboratory of Science and Technology on Advanced Composites in Special Environments, Harbin Institute of Technology; 92 Xidazhi Street, Harbin, 150001, PR China\\
$^{b}$ \quad Institut FEMTO-ST, CNRS, Universit{\'e} Bourgogne Franche-Comt{\'e}, 25000 Besan\c{c}on, France}
\fntext[myfootnote]{tanhf@hit.edu.cn}

\begin{abstract}
Cork is a natural amorphous material with near-zero Poisson's ratio that is ubiquitously used for sealing glass bottles.
It is an anisotropic, transversally isotropic, composite that can hardly be scaled down. 
Here, we propose a new class of isotropic and reusable cork-like metamaterial that is designed from an hybrid truss-lattice material to show an isotropic Poisson's ratio close to zero.
Optimization is conducted using a multi-objective genetic algorithm, assisted by an elliptical basis function neural network, and coupled with finite element simulations.
The optimal micro-structured metamaterial, fabricated by two-photon lithography with a lattice constant of 300 \micro\meter, has an almost isotropic Poisson's ratio smaller than 0.08 in all directions.
It can recover $96.6\%$ of its original shape after a compressional test exceeding $20\%$ strain.
\end{abstract}


\end{frontmatter}


\section{Introduction}

Poisson's ratio $\upsilon$ is defined as the negative ratio of transverse to longitudinal strain \cite{timoshenko1970}. 
For a stable, isotropic and linear elastic material, Poisson's ratio is bound to remain between $-1$ \cite{milton1995elasticity,huang2016pentamodal}, corresponding to ’dilational’ or auxetic materials, and $0.5$, a limit defining the ’incompressible’ solid set by a positive energy requirement \cite{sokolnikoff1956mathematical,gercek2007poisson}.
In nature, most conventional isotropic materials have a positive Poisson's ratio.
Rubber, as well as most liquids, exhibits a Poisson's ratio of nearly $0.5$. 
Rigid metals and polymers as a rule have a poisson's ratio ranging between $0.2$ and $0.45$ \cite{milton1995elasticity,greaves2011poisson}.
For other soft metals and polymers, Poisson's ratio is usually between $0.33$ and $0.5$. 
By contrast, only a few natural materials such as bone have negative Poisson's ratio \cite{wojciechowski2015auxetics}.

Recent advances in topological structural design have enabled the enlargement of the family of isotropic auxetics \cite{buckmann2014three}.
Carta \textit{et al.} utilized threefold symmetry of the arrangement of voids to design a two-dimensional porous isotropic auxetic solid \cite{carta2016design}.
By embedding random re-entrant inclusions into a matrix, Hou \textit{et al.} developed 2D composite structures with isotropic negative Poisson's ratio \cite{hou2012novel}. 
Combining the symmetry of a cubic lattice and that of additional diagonal elements, Cabras \textit{et al.} presented a class of pin-jointed auxetic three-dimensional isotropic lattice  material \cite{cabras2016class}. 
Furthermore, by adopting finite small connections, B\"{u}ckmann \textit{et al.} designed, fabricated and characterized  a three-dimensional auxetic isotropic metamaterial reaching an ultimate Poisson's ratio of $-0.8$ \cite{buckmann2014three}. 
Lately, Frenzel et al. used auxetics combined with chirality to observe acoustical activity
\cite{frenzel2017three,frenzel2019ultrasound}.  

Isotropic structural materials with positive Poisson's ratio are generally designed for bearing different types of mechanical loads \cite{messner2016optimal,gurtner2014stiffest,berger2017mechanical,tancogne20183d} or absorbing energy \cite{bonatti2019mechanical}.
The most popular way to optimise isotropy is to overlapp different structures in order to increase the number of equivalent directions and thus, via geometry increase, isotropy \cite{tancogne2018elastically1,tancogne2018elastically,gurtner2014stiffest,xu2016design,latture2018design}.
Gurtner \textit{et al.} proposed the first optimal and isotropic three-dimensional truss-lattice structure \cite{gurtner2014stiffest}.
Tancogne \textit{et al.} further formulated analytical conditions on the lattice topology to achieve elastic isotropy \cite{tancogne2018elastically1} and studied the effect of bending ratio to axial stiffness of the micro-strut on structural isotropy \cite{tancogne2018elastically}.
Bonatti \textit{et al.} recently reported a family of elastically-isotropic shell-lattice materials whose Young’s modulus is always higher than that of optimal isotropic truss-lattices and approaches the Hashin--Shtrikman bound at high relative densities \cite{bonatti2019mechanical}.  
Berger \textit{et al.} presented a class of cubic-octet hybrid closed foams achieving the Hashin--Shtrikman upper bounds on isotropic elastic stiffness \cite{berger2017mechanical}.
Tancogne \textit{et al.} identified a class of low-density plate-lattice metamaterial showing optimal isotropic stiffness and nearly isotropic yield strength \cite{tancogne20183d}.

\begin{figure*}[!bh]
\centering
\includegraphics[width=14cm]{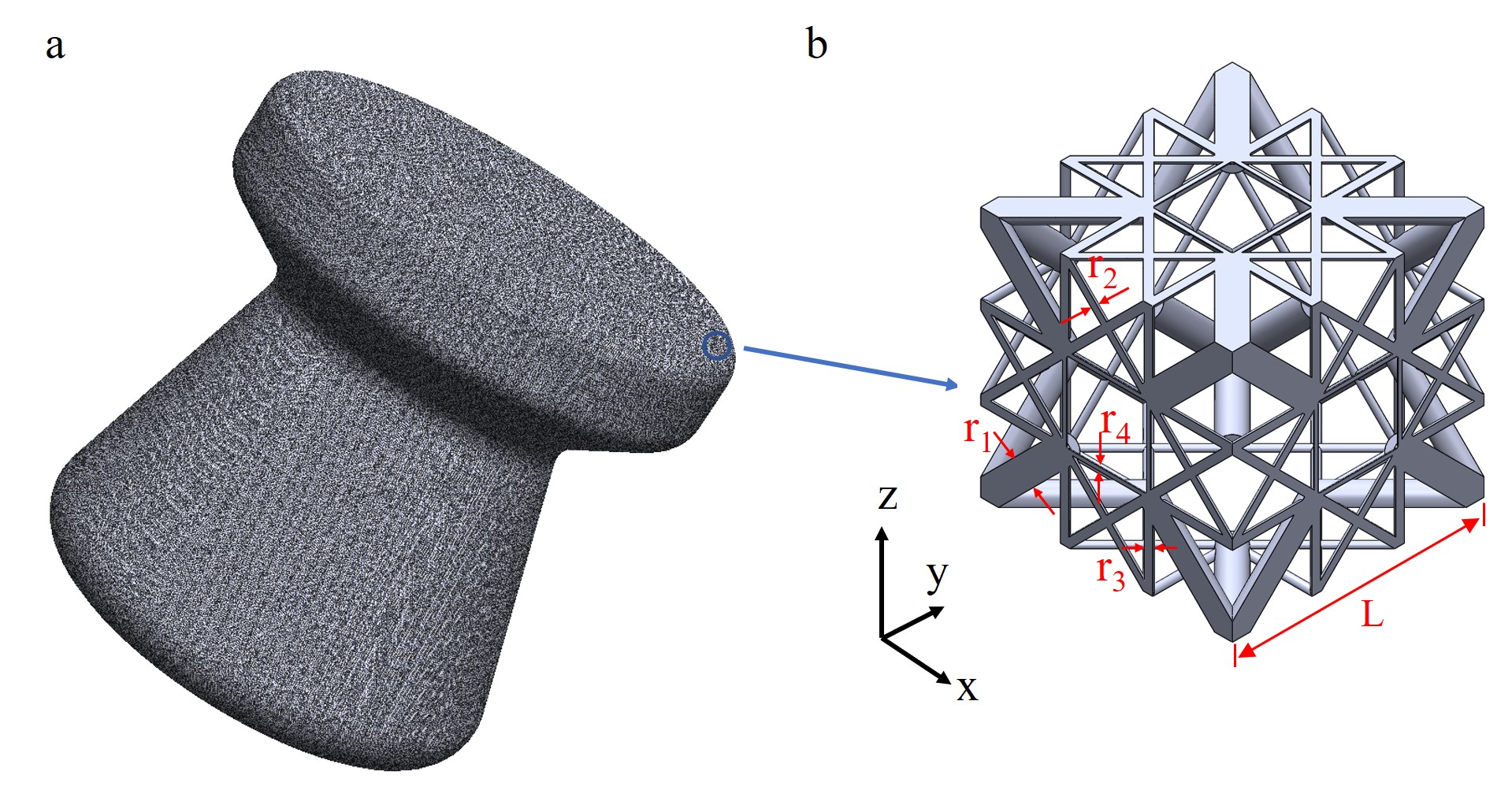}
\caption{Principle of the truss lattice material with near-zero Poisson's ratio.
(a) Artistic illustration of a truss lattice bottle stopper and (b) corresponding representive unit cell with geometrical parameters indicated.   
}
\label{fig1}
\end{figure*}

Cork, a conventional natural material, is emblematic among near-zero Poisson's ratio materials \cite{gibson1981structure,fortes1989poison,stavroulakis2005auxetic}.
It shows very little lateral expansion when compressed and is widely used to seal bottles, especially for wine.  
As a composite, it is almost transversally isotropic and Poisson's ratio is indeed a symmetric  tensor.
Independent Poisson's ratio constants $\upsilon_{12} = 0.097$, $\upsilon_{13}= 0.064$, and $\upsilon_{23}=0.26$ have been reported for cork \cite{fortes1989poison}.
Recently, some efforts were made to design isotropic zero Poisson's ratio materials.  
Based on truss or thin frame beam theory, Sigmund presented a three dimensional optimal structure with zero poisson's ratio \cite{sigmund1995tailoring}. 
Starting from a different structure, Guth \textit{et al.} proposed another kind of 3D pin-jointed structure \cite{guth2015optimization}. % That sentence is difficult to understand...
However, to the best of our knowledge, no cork-like isotropic structure has been validated experimentaly thus far. 
Moreover, subject to limitations of numerical algorithms, the effect of the nodal overlapping volume was not considered, which we find seriously influence mechanical properties, including isotropy and Poisson's ratio.   

In this paper, we aim at designing an isotropic near-zero Poisson's ratio material based on a periodic microstructure with cubic symmetry, that can be scaled easily and fabricated additively.
We base our design on the hybrid truss lattice structure of Fig.~\ref{fig1} that was first presented by Sigmund \cite{sigmund1995tailoring}.
The unit cell follows simple cubic symmetry.
Isotropy and near-zero Poisson's ratio are set as goals of a multi-objective optimization procedure where the radii of the struts are the optimized parameters.
Optimization results in an almost isotropic design with Poisson's ratio less than $0.08$ in all directions.
Samples are printed using two-photon polymerization at a lattice constant of 300 \micro\meter~in two different crystallographic directions, [100] and [110].
Uniaxial compression tests confirm the isotropic near-zero Poisson's ratio but also the recovery of the material after enduring strains up to $20\%$.

\section{Evaluation of isotropy and Poisson's ratio}

The constitutive law of linear elasticity of three-dimensional composites relates the stress tensor $\bm{\sigma}$ to the strain tensor $\bm{\epsilon}$ via an effective order-4 symmetric stiffness tensor $\bm{C}$ as
\begin{equation}
\centering
\bm{\sigma}=\bm{C}: \bm{\epsilon},
\end{equation}
where $C_{ijkl}=C_{klij}=C_{jilk}$.
For lattice materials with simple-cubic symmetry \cite{rand2007analytical,buckmann2014three}, the effective stiffness tensor has only three independent elements and can be rewritten in Voigt notation \cite {voigt1910lehrbuch},
\begin{equation}
\bm{C}
=\begin{bmatrix}
C_{11} &  C_{12}  & C_{12}& 0 & 0 & 0\\
& C_{11}   & C_{12} &  0 & 0 & 0\\
& & C_{11} &  0 & 0 & 0\\
& & &  C_{44} & 0& 0\\
&  sym  & & & C_{44} & 0 \\
& & & & & C_{44}
\end{bmatrix} .
\end{equation}

Using the Christofell equation for elastic waves \cite{laudeBOOK2015,christensen_kadic_kraft_wegener_2015}, the independent stiffness elements can be expressed using the effective mass density and phase velocities in selected directions of propagation.
The effective mass density $\rho$ is defined as the product of volume filling fraction $f$ by the mass density $\rho_0$ of the constituent material \cite{gibson1999cellular}.
Only three phase velocities $v$ are required to identify all three independent stiffness constants.
We consider the three bulk waves in direction $[110]$.
One is a pure-shear wave S1 polarized along direction $[001]$, the other two are quasi-longitudinal L and quasi-shear S2 waves with mixed polarization in the $(x,y)$ plane.
For propagation in direction $[110]$, the Christofell equation leads to \cite{tsang1983sound,buckmann2014three}
\begin{align}
C_{44}&=\rho v_{\rm{S}1}^2, \label{stiffness44} \\
C_{12}&=\rho v_{\rm{L}}^2 - \rho v_{\rm{S}1}^2 - \rho v_{\rm{S}2}^2, \label{stiffness12} \\
C_{11}&=\rho v_{\rm{L}}^2 - \rho v_{\rm{S}1}^2 + \rho v_{\rm{S}2}^2. \label{stiffness11}
\end{align} 
For propagation along direction $[100]$, Eq. \eqref{stiffness44} would be unchanged whereas Eq. \eqref{stiffness11} would give $C_{11}=\rho v_{L}^2$.
Isotropy requires velocity to be independent of the direction of propagation and hence implies
\begin{equation}
v_{\rm{S}1} = v_{\rm{S}2} \mathrm{~~along~direction~} [110]. \label{1}
\end{equation}
Reciprocally, if Eq. \eqref{1} holds then there are only two independent stiffness constants instead of three and the stiffness tensor is isotropic.
As a whole, Eq. \eqref{1} is a necessary and sufficient condition for isotropy.
Poisson's ratio for compression along the principal axes can be expressed as \cite{hill1952elastic,bower2009applied}
\begin{equation}
\upsilon=\frac{C_{12}}{C_{11}+C_{12}}. \label{upsilon}
\end{equation} 
Hence, we can estimate Poisson's ratio in direction $[110]$ using the following formula
\begin{equation}
\upsilon = \frac{v_{\rm{L}}^2 - v_{\rm{S}1}^2 - v_{\rm{S}2}^2}{2(v_{\rm{L}}^2 - v_{\rm{S}1}^2)} , \label{2}
\end{equation}
where velocities are measured along direction $[110]$.
If isotropy is simultaneouly achieved, formula \eqref{2} is valid for all directions of propagation.

In practice, velocities are obtained numerically using a finite element model of the unit cell in Fig. \ref{fig1}(b) subjected to Bloch periodic boundary conditions.
A small wavenumber $k = \pi / (100 L)$ is considered along direction $[110]$ and eigenfrequencies are obtained.
The three lowest eigenfrequencies, when divided by $k$, give velocities $v_{\rm{S}1}$, $v_{\rm{S}2}$ and $v_{\rm{L}}$; they are readily classified as longitudinal or shear by comparing the polarization of the eigenfunctions.

We note another useful expression for the Poisson's ratio for cubic symmetry that is valid for an arbitrary compression direction~\cite{wojciechowski2005poisson,paszkiewicz2001unified,buckmann2014three}. 
\begin{equation}
\upsilon(\phi,\theta) = -\frac{Ar_{12}+B(r_{44}-2)}{16[C+D(2r_{12}+r_{44})]}
\label{Eq9}
\end{equation}    
with
\begin{align}
r_{12} &= \frac{S_{12}}{S_{11}},\\
r_{44} &= \frac{S_{44}}{S_{11}},\\
A &= 2[53+4\cos(2\theta)+7\cos(4\theta) \nonumber \\ &+ 8\cos(4\phi)\sin^{4}(\theta)], \\
B &= -11+4\cos(2\theta)+7\cos(4\theta) \nonumber \\ &+ 8\cos(4\phi)\sin^{4}(\theta),\\
C &= 8\cos^{4}(\theta)+6\sin^{4}(\theta) \nonumber \\ &+ 2\cos(4\phi)\sin^{4}(\theta),\\
D &= 2 [\sin^{2}(2\theta) + \sin^{4}(\theta) + \sin^{4}(2\phi)] ,
\end{align}  
where $(\theta, \phi)$ are the azimuthal and polar angles in spherical coordinates.
The compliance tensor $\bm{S}$ is the inverse of the stiffness tensor $\bm{C}$.

%%%%%%%%%%%%%
\section{Optimization of the structure}

\subsection{Optimization strategy}

The cubic-symmetry truss lattice structure of Fig.~\ref{fig1} was selected for optimization.
The corresponding representive unit cell model contains 64 struts of four different types.
The unit cell length $L$ being fixed to 300\,\micro\meter, there are four geometrical parameters, $(r_{1}, r_{2}, r_{3}, r_{4})$, available for optimization.
The ranges of the design parameters were fixed as $14\,\micro\meter\leq r_{1} \leq 16\,\micro\meter$, $4\,\micro\meter \leq r_{2} \leq 6\,\micro\meter$,  $4\,\micro\meter \leq r_{3} \leq 6\,\micro\meter$, and $2\,\micro\meter \leq r_{4} \leq 4\,\micro\meter$.
Compared with the structure originally proposed by Sigmund \cite{sigmund1995tailoring}, we consider larger values for $r_{1}$ but smaller values for $r_{2}$.

The optimization problem aims at simultaneously imposing the isotropy condition \eqref{1} and minimizing Poisson's ratio \eqref{2}.
The objective function to be minimized is thus selected as
\begin{align}
E(r_{1}, r_{2}, r_{3}, r_{4}) &= 2 \frac{| v_{\rm{S}1} - v_{\rm{S}2}|}{v_{\rm{S}1} + v_{\rm{S}2}} \nonumber \\
&+ \left| \frac{v_{\rm{L}}^2 - v_{\rm{S}1}^2 - v_{\rm{S}2}^2}{2(v_{\rm{L}}^2 - v_{\rm{S}1}^2)} \right|.
\label{objective}
\end{align}
This objective function uses equal weighting factors for both criteria, resulting in a balanced multiobjective optimization problem.

A commercial finite element software package (COMSOL Multiphysics) was adopted to caculate the required velocities. 
The truss lattice structures were modeled with several hundred of thousands of linear tetrahedral finite elements.
The constituent material chosen is assumed isotropic and linearly elastic with Young's modulus $E_0=2$~GPa, $\upsilon_0=0.4$, and mass density $\rho_0 = 1000$~kg$\cdot$m$^{-3}$.

\begin{figure}[!bt]
	\centering
	\includegraphics[width=9cm]{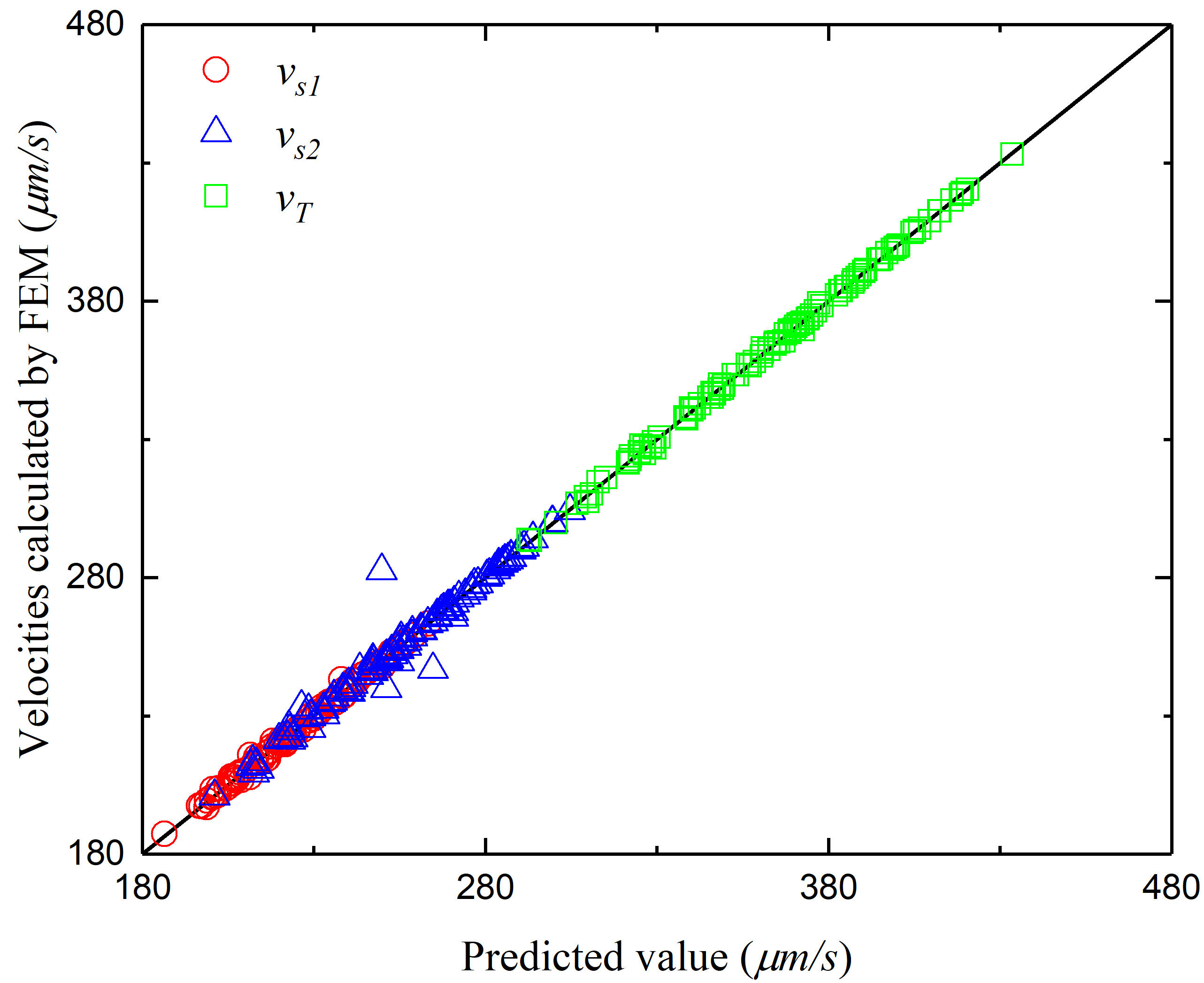}
	\caption{Comparison of velocities predicted by EBFNN with velocities obtained by FEM. 
	}
	\label{fig2}
\end{figure}

The parameter space was sampled in order to reduce the computational burden during optimization.
Toward this end, a surrogate model was created from a finite number of parameter space samples.
One hundred sample points were first generated according to optimal Latin-hypercube design (OLD).
This method was used to distribute sample points so that they are well spread over the design region without replicated coordinate values, often symmetric, and nearly optimal \cite{park1994optimal}.  
The generated sample points are listed in Table S1 of the Supplemental Material.
A surrogate model was then generated and optimization was performed on the reduced parameter space, as described next.

\subsection{Surrogate models}  

The elliptical basis function neural network (EBFNN) technique has proven effective in approximating a continuous function of $n$ variables in very complex cases \cite{bishop1991improving,schilling2001approximation,mak2000estimation}.
From the parameter space samples, a EBFNN was constructed to generate approximate surrogate models of the three velocities $v_{S1}$, $v_{S2}$ and $v_{L}$. 

\begin{table}[!bt]
	\captionsetup{singlelinecheck=off, skip=4pt}
	\caption{Accuracy measures of the EBFNN surrogate models.}
	\centering
	\begin{tabular}{lll}
		\toprule[2pt]
		Velocity & RMSE & $R^{2}$ \\
		\midrule
		$v_{S1}$   & 0.01546 & 0.99534 \\
		$v_{S2}$   & 0.0403  & 0.96955 \\
		$v_{L}$    & 0.00591 & 0.99934 \\
		\bottomrule[2pt]
	\end{tabular}
    \label{tab1}
\end{table}

The coefficient of determination ($R^{2}$) and the root mean square error (RMSE) are used to evalute the reliability of the surrogate models. 
These estimators are defined as
\begin{align}
R^{2} &= 1-\frac{\displaystyle \sum_{i=1}^{n}(y_{i}-\hat{y_{i}})^2}{\displaystyle \sum_{i=1}^{n}(y_{i}-\bar{y})^2} ,  \\
\mathrm{RMSE} &= \sqrt{ \frac{1}{n} \sum_{i=1}^{n} (y_{i}-\hat{y_{i}})^2} .
\end{align} 
In these expressions, $n$ is the number of samples, $y_{i}$ are the actual values of objective function at the sample points, $\hat{y_{i}}$ are the values predicted by the objective function, $\bar{y}$ is the mean value of objective function over all sample points.
All sample points defined by OLD are used for cross-validation error analysis.
The closer $R^{2}$ is to 1 and RMSE is to 0, the more accurate the model. 
For all surrogate models, $R^{2}$ is larger than 0.969 and RMSE is smaller than $4\%$, as listed in Table \ref{tab1}.
These values indicate that the surrogate models have high credibility. 
Fig. \ref{fig2} compares the velocities predicted by the surrogate models with the actual velocities, for all sample points.
It can also be observed that the prediction error remains small in all cases.
Of course, the usefulness of the surrogate models is to produce smooth estimates of the velocities for any continuous value of the quadruplet $(r_{1}, r_{2}, r_{3}, r_{4})$.

\subsection{Optimization} 

\begin{table*}[!tb]
\captionsetup{singlelinecheck=off, skip=4pt}
\caption{Optimization results. Geometrical parameters, angular velocities in the $[110)$ direction, and minimal and maximal values of Poisson's ratio $\upsilon$ for all compression directions are given for the initial and selected optimized designs. }
\begin{tabular*}{\hsize}{@{}@{\extracolsep{\fill}}lllllllllllll@{}}
\toprule[2pt]
	structure & $r_{1}$ (\micro\meter)  & $r_{2}$ (\micro\meter) & $r_{3}$ (\micro\meter) & $r_{4}$ (\micro\meter) & $v_{s1}$ (\micro\meter/s) & $v_{s2}$ (\micro\meter/s) & $v_{L}$  (\micro\meter/s) & $\upsilon_{min}$ & $\upsilon_{max}$ \\
	\midrule
	Initial & 14.444  & 4.040 & 5.111 & 3.677 & 207.483 & 283.900 & 383.421 & 0.112 & 0.239 \\
	Optimum 1  & 15.000  & 4.500 & 5.100 & 2.400 & 218.590 & 219.256 & 323.220 & 0.076 & 0.077\\
    Optimum 2  & 15.960  & 4.707 & 4.303 & 2.485 & 211.481 & 211.703 & 314.556 & 0.086 & 0.087 \\
	Optimum 3  & 15.535  & 4.121 & 4.909 & 2.222 & 198.819 & 201.040 & 293.675 & 0.067 & 0.073\\
	Optimum 4  & 15.000  & 4.300 & 4.850 & 2.350 & 211.037 & 213.036 & 313.001 & 0.075 & 0.080\\
	\bottomrule[2pt]
	\end{tabular*}
	\label{tab2}
\end{table*}

\begin{figure}[!b]
\centering
\includegraphics[width=9cm]{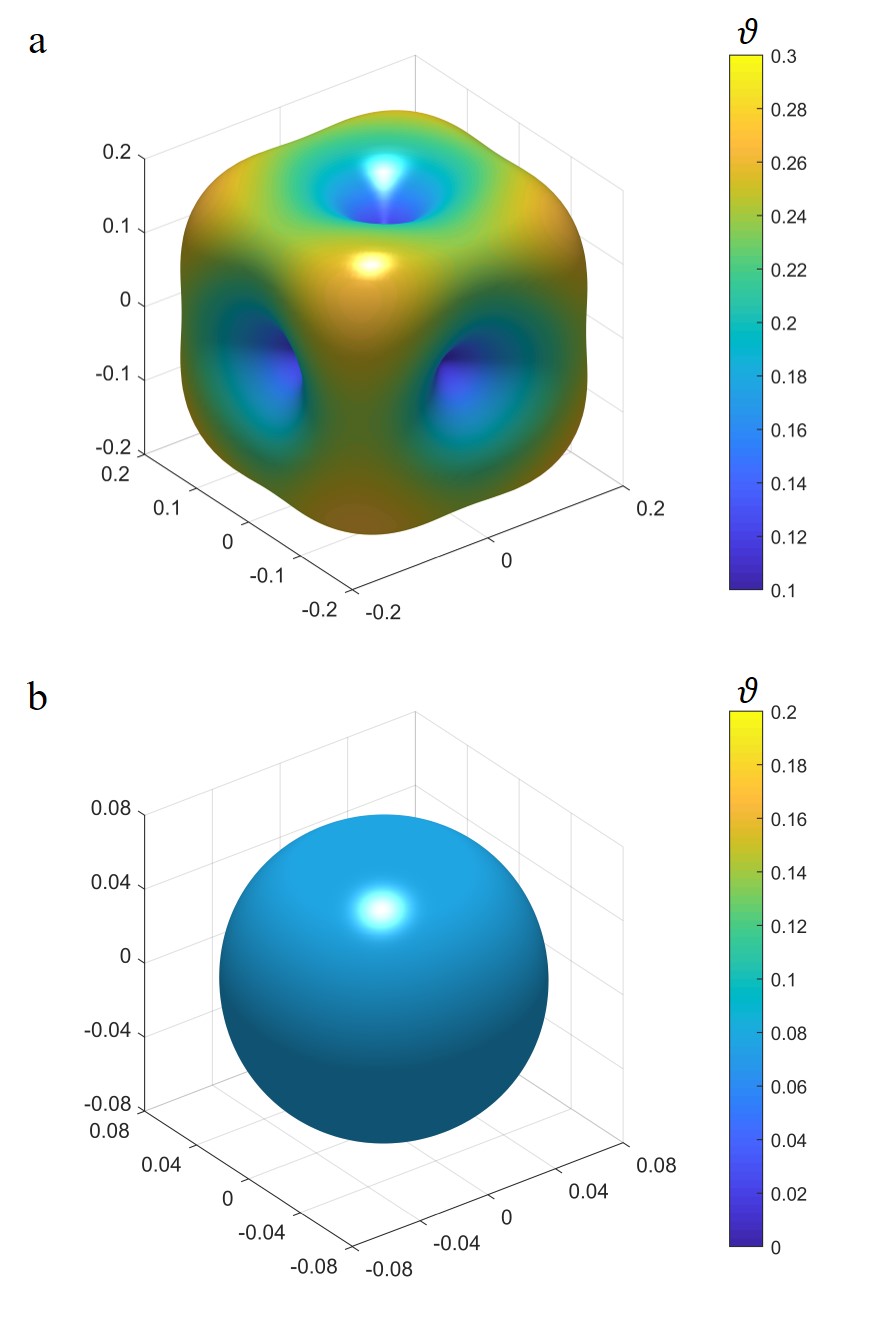}
\caption{Three dimensional polar plot of the Poisson's ratio following by Eq. (\ref{Eq9}) for (a) the initial structure and (b) the optimal isotropic structure 1.}
\label{fig3}
\end{figure}
 
Non-dominated sorting genetic algorithm (NSGA-II) \cite{deb2002fast} is used to find solutions to the optimization problem. 
The population size is $12$ and the number of generations is $2000$.
Distribution indexes are $\eta_{c}=20$ for crossover and $\eta_{m}=10$ for mutation.  
The probability of crossover is $0.9$.
To account for possible errors caused by the surrogate models, not only the optimum solution but also some local minima were extracted.
By comparing simulations and optimization results, we picked up the four optimum designs listed in Table \ref{tab2}.
Velocities and Poisson's ratios are estimated by conducting finite element simulations again after optimization.

The initial structure was quite anisotropic, with the Poisson's ratio varying between $0.112$ and $0.239$. 
Moreover, the minimum Poisson's ratio was larger than the upper bound for cork, $0.1$.
After optimization, an almost isotropic value $\upsilon \approx 0.08$ is obtained for the four selected designs.
Fig. \ref{fig3} plots the Poisson's ratio in spherical coordinates for both the initial and the optimum structure 1.
The response of the optimum structure is clearly much more isotropic than cork.

%%%%%%%%%%%%%%%%%%%%%%%%%%%%%%%%%
\section{Experiment}

\begin{figure*}[!bt]
\centering
\includegraphics[width=13cm]{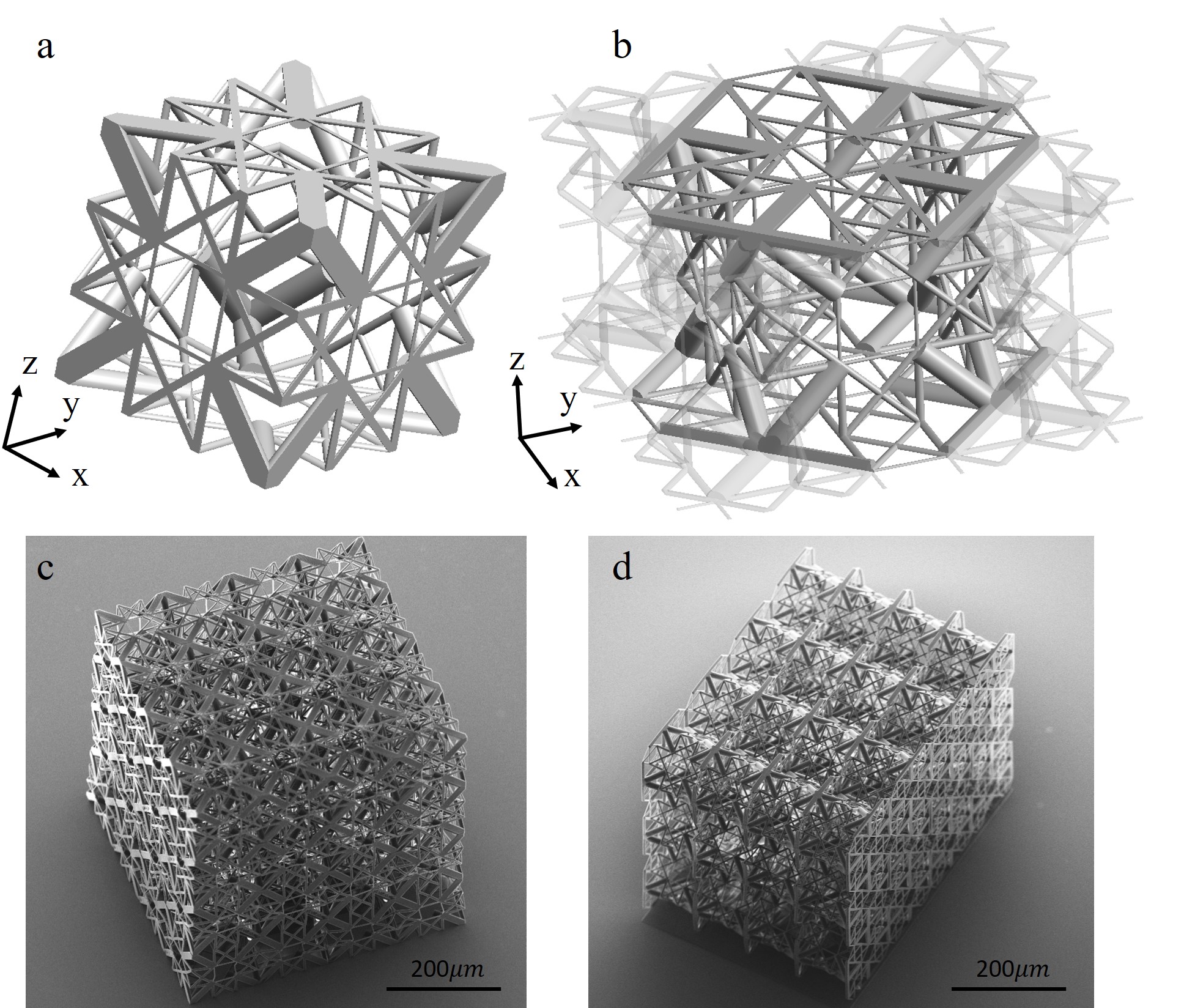}
\caption{
Unit cell models of the isotropic truss lattice material for (a) the $[100]$ direction and (b) the $[110]$ direction. Electron micrographs are shown for (c) the $[100]$ fabricated sample with $4 \times 4 \times 4$ unit cells and (d) the $[110]$ fabricated sample with $4 \times 4 \times 3$ unit cells. 
}
\label{fig4}
\end{figure*} 

\begin{figure*}[!tbh]
\centering
\includegraphics[width=13cm]{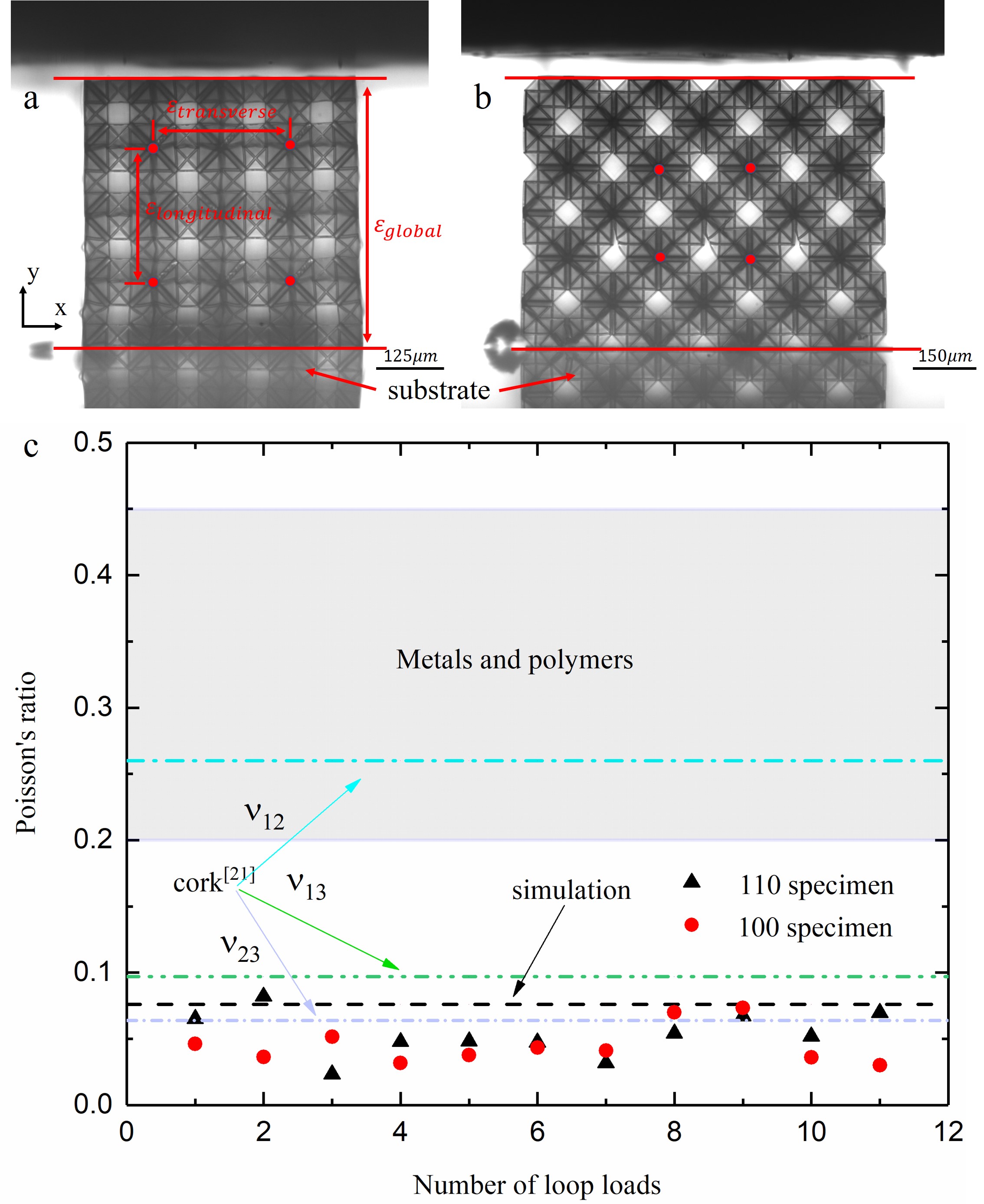}
\caption{
Definition of reference points and reference lines used to determine the transverse strain, the longitudinal strain and the global strain, for (a) the $[100]$ sample and (b) the $[110]$ sample.
(c) Poisson's ratio of the samples is plotted as a function of the number of experimental loop loads.
Values for the FEM simulation, cork, metals and polymers are shown for comparison.
}
\label{fig5}
\end{figure*}   

All experimental samples are made from the 'IP-Dip' resin using the commercially available laser lithography system Photonic Professional GT (Nanoscribe GmbH, Germany).
A drop of a negative-tone photoresist is placed on top of a fused silica substrate ($25 \times 25 \times 0.7$ \milli\cubic\meter) and polymerized using a femtosecond pulsed laser with vacuum wavelength $ \lambda = 780$ \nano\meter.
The laser beam is focused by using a dip-in $\times$63 objective lens with 1.4 numerical aperture. 
A Galvanometric scan speed of 10 \meter\per\second~ was used for the whole fabrication process.
After polymerization is achieved, the sample is developed in a PGMEA (1-methoxy-2-propanol actetate) for 20 minutes to remove the unexposed photoresist.

Two different crystallographic directions are considered, $[100]$ and $[110]$.
Fig. \ref{fig4} shows the unit cell models and the corresponding additively manufactured samples.
The $[100]$ sample, which is composed of 4$\times$4$\times$4 unit cells, is constructed by stacking the corresponding unit cell in the three principal directions.
Noting that the Poisson's ratio of lattice materials is mainly affected by the aspect ratio of  micro-struts rather than by other geometrical parameters \cite{buckmann2014three}, we adopted the aspect ratios obtained from optimization and scaled the unit cell length proportionaly.
The detailed geometrical parameters are: $L=125$ \micro\meter, $r_{1}=6.3$ \micro\meter, $r_{2}=1.9$ \micro\meter, $r_{3}=2.1$ \micro\meter, and $r_{4}=1$ \micro\meter.
  
The $[110]$ sample is generated by cutting out a $[100]$ structure $2 \times 2 \times 1$ along the vertical direction. 
The horizontal basis vectors are then along directions $[110]$ and $[\bar{1}10]$.
It should be noted that the geometrical features of the $[110]$ unit cell can be described by that of the corresponding $[100]$ unit cell.
Here, geometical parameters are $L=150$ \micro\meter, $r_{1}=7.6$ \micro\meter, $r_{2}=2.3$ \micro\meter, $r_{3}=2.5$ \micro\meter, and $r_{4}=1.2$ \micro\meter.
The $[110]$ sample contains $4 \times 3 \times 4$ unit cells.
The external dimensions are $848.4$ \micro\meter~ $\times$ $636.3$ \micro\meter~ $\times$ $450$ \micro\meter.

As shown in Fig. S1 of the supporting material, the samples are placed between a fixed glass substrate and a flat loading device.
The loading device is driven by a stepping motor with an attached force sensor.
Position is directly read from the linear stage.
The position is only used to monitor the fatigue of the material.
The true strain is obtained via image cross correlation.
To test the recovery ability of the samples, repeated compressive experiments are carried out at a speed of $0.001$ \milli\meter\per\second, during which the applied displacement increases with loop number.
A digital camera equiped with a $20\times$ objective lens facing the sample is used to monitor the deformation of the lateral faces and hence to measure Poisson's ratio.
Digital image correlation \cite{mak2000estimation} is used to track and analyze the displacement with sub-pixel resolution.
To reduce the influence of boundaries, Poisson's ratio is calculated from the average local strain and the average transverse strain measured from 4 reference circles at the central row of unit cells as depicted in Fig. \ref{fig5}.  
Global strain is determined by measuring the distance between the reference lines.

\begin{figure*}[!tbh]
\centering
\includegraphics[width=13cm]{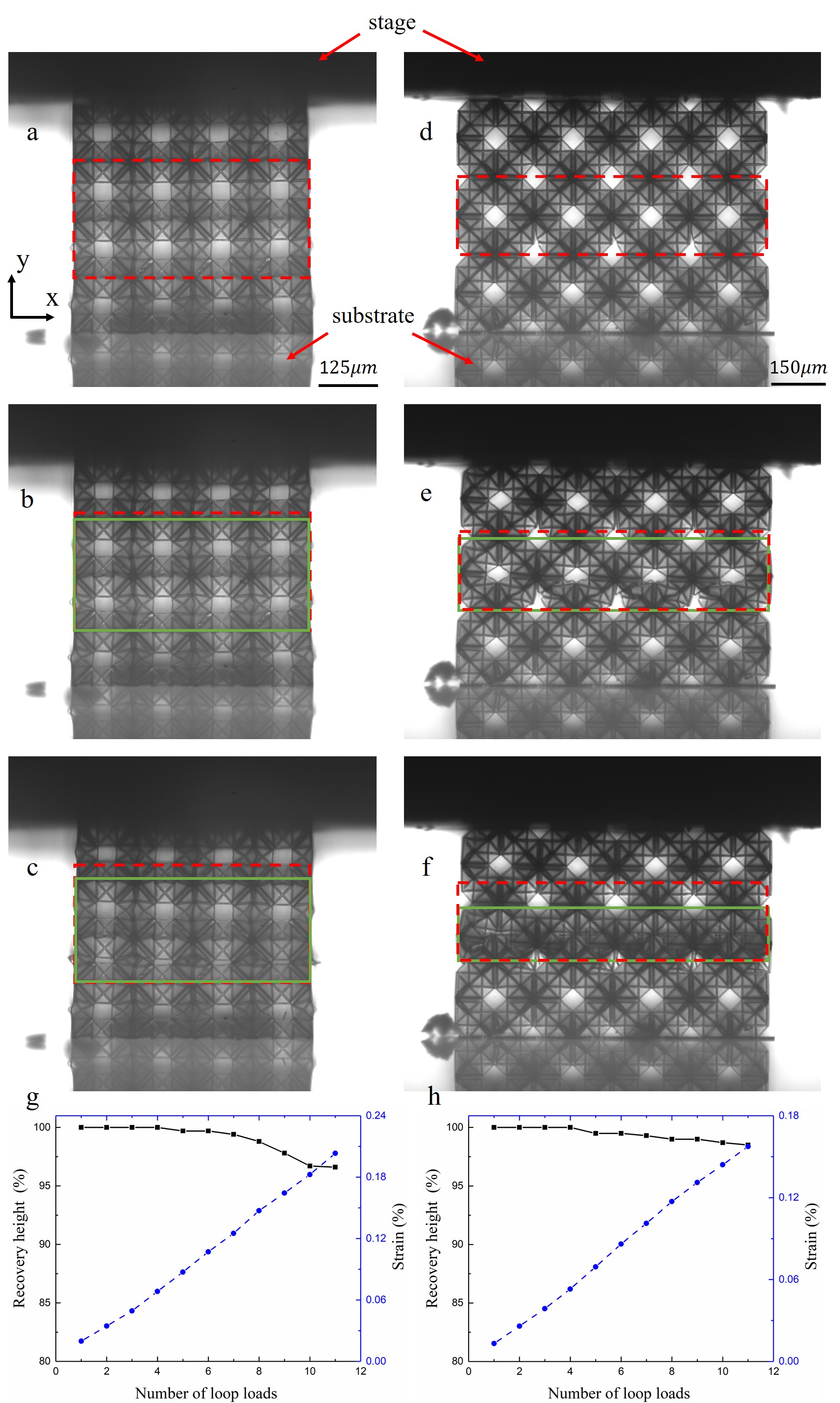}
\caption{(a-c) Views of the deformed $[100]$ sample at $0\%$, $5\%$ and $10\%$ strain. (d-f) Views of the deformed $[110]$ sample at $0\%$, $5\%$ and $10\%$ strain.
The red dashed square and the green solid square are the initial and the deformed shapes of samples, respetively.
(g,h) Recovery ability of the [100] and the [110] samples and maximum applied strain as a function of the loop number.
}
\label{fig6}
\end{figure*}

Fig. \ref{fig5}(c) presents the measured Poisson's ratio of the [100] sample and the [110] sample.
For both samples, experimental data are in fair agreement with simulation results of Table \ref{tab2}.
The measurements are generally found to be smaller than the computed value.
The contrast between samples shows that the proposed structure has a more isotropic response than cork.
Moreover, the number of loop loading has a limited impact on the value of the Poisson's ratio.
Even though some micro-struts break at large applied strain, the measured initial Poisson's ratio always fluctuates around the designed value.
 
Fig. \ref{fig6} summarizes the results of eleven cyclic compression experiments.
%{\color{red} Red dashed square and green solid square are undeformable and deformable shapes for central rows of samples with different applied strain.
A large vertical deformation together with a very small horizontal deformation are observed under compression, indicating that the structural materials have a nearly zero global Poisson's ratio.
For both samples, the maximum applied strain increases almost linearly with the loop number. 
During the first and the last loop, the maximum strains of the $[100]$ sample are $2\%$ and $20\%$, respectively. 
As long as the applied strain remains smaller than $7\%$, the sample can recover completely after unloading. 
This property may be attributed to elastic buclking of the slender members in the micro-lattice.
When the applied strain is increased above $7\%$, however, the recovery ability of sample weakens slightly.
With a maximum applied strain of $20\%$, the sample can still recover almost $96.6\%$ of its original height.     
In principle, the samples should possess even better recovery ability and should withstand larger strains.
However, the slender micro-struts are very sensitive to flaws and imperfections.
Hence the deformation of the sample may not be homogeneous and failure may start within any layer in the fashion of brittle break of the micro-structs. 
The compressive experiment validates our hypothesis (see SM. (1-6)).
A similar trend regarding the recovery ability is found for the [110] sample.  
At large strain, brittle break of micro-struts is also the dominating failure mode of the tested sample. 
The only difference is that the recovery ability is further weakening.
The $[110]$ sample seems to be even more sensitive to flaws than the $[100]$ sample.
With a maximum applied strain of $16\%$, the $[110]$ sample can almost recover $98.5\%$ of its original height.

%%%%%%%%%%%%%%%%%%%%%%%%%%%
\section{Conclusion}

A new class of isotropic reusable cork-like metamaterial with near-zero Poisson's ratio was designed using a multi-objective genetic algorithm assisted by an elliptical basis function neural network combined with finite element simulations.
We derived an objective function for simultaneously imposing elastic isotropy and controlling the value of Poisson's ratio.
The optimal structures were fabricated and tested under repeated compression experiments.
Results show that the samples fabricated using two-photon lithography have an almost isotropic near-zero Poisson's ratio.
Furthermore, they can almost recover $96.6\%$ of their original shape after the eleventh compressional test exceeding $20\%$ strain.
The number of loop loadings has a limited impact on the value of Poisson's ratio.
Even though some micro-structs break at large applied strain, the Poisson's ratio still fluctuates around the designed value.  

\bigskip
\section*{Acknowledgments}

This work was supported by Foreign Short-term Visiting Program for Doctoral Students at HIT.
N.L., S.M., J.M., M.K. and V.L. acknowledge support by the EIPHI Graduate School (contract ”ANR-17-EURE-0002”) and the  French Investissements d’Avenir program, project ISITE--BFC (contract ANR-15-IDEX-03).
This work was partly supported by the french RENATECH network and its FEMTO-ST technological facility.

\bibliographystyle{elsarticle-num} 
\bibliography{paper}


\end{document}